# Enhancing the magnetism and giant anomalous Hall effect in thin Fe-Al films via B2 nanophase growth

D. A. Tatarskiy[1], A. A. Nazarov[2], Y. M. Kuznetsov[1], A. V. Zdoroveyshchev[1], I. Y. Pashenkin[2], P. A. Yunin[1,2], S. A. Churin[2], E. S. Demidov[1], M. V. Sapozhnikov[1,2], N. I. Polushkin[2]

[1] *Lobachevsky State University of Nizhny Novgorod, Nizhny Novgorod 603950, Russia*
[2]*Institute for Physics of Microstructures, Russian Academy of Sciences, Nizhny Novgorod GSP-105, Russia*

**ABSTRACT**. The properties of alloys that undergo to chemical order-disorder transformations depend heavily on the degree of ordering in the crystal lattice. In the literature, it is well established that the ordering in a magnetic alloy such as Fe-rich ($x$>0.5) $Fe_xAl_{1-x}$ leads to reducing its magnetization and even to a transition from the ferromagnetic (FM) to paramagnetic (PM) state at $x$<0.7. Studying the ordering kinetics in thin (50 nm) $Fe_xAl_{1-x}$ films with a non-stoichiometric composition (0.5<$x$<0.7), we demonstrate the opposite behavior: When the alloy is aged at a high temperature $T_a$>600 °C, the ordering process is accompanied by an increase in magnetization and related properties. For example, we find the further enhancement of the giant anomalous Hall (AH) effect found recently in $Fe_xAl_{1-x}$ alloys. Based on both experimental data and theoretical modeling, we argue that these properties are enhanced due to the nucleation and growth of the B2-$Fe_{0.5}Al_{0.5}$ phase. Growing B2 nanocrystals enable segregation and clustering of excess Fe in the alloy. It has been revealed that the PM phase, which is formed in the aged samples and contains Fe-enriched superparamagnetic clusters, contributes to the AH resistivity even more than the FM phase in the as-grown sample. Our findings open a route for improving the properties of functional alloys.

## Introduction

The study of the ordered and disordered states in Fe(Co)-rich $Fe(Co)_xAl_{1-x}$ alloys and their derivatives, such as $Fe_2(Co_2)XAl(Si)$ Heusler alloys with X = Co, Fe, Mn, Cr, V,.., is currently of great interest due to their half-metallic behavior [1, 2] and sharp magnetic transitions [3, 4], which have thermoelectric [5, 6], spintronic [7-13], magnetocaloric [14-16], and microbiological [17] applications. The order-disorder transformations in binary $Fe_xAl_{1-x}$ alloys, which crystallize into the cubic lattice of the CsCl type, have been studied for decades, since the first observations of ordering and superstructure formation in these systems in 1932 [18]. In a perfectly chemically ordered stoichiometric alloy, atoms of all species occupy regular positions in the atomic lattice, and so, the long-range order (LRO) in the alloy becomes equal to unity. The vast majority of alloys exhibit a strong dependence on their physical properties on the atomic arrangement and LRO. In particular, studies on the magnetic and structural properties of the $Fe_xAl_{1-x}$ system have shown [19-29] that magnetization in the alloy only decreases with increasing LRO. In Figure 1a, this magnetization behavior is shown schematically as a function of aging time. In the disordered A2 state, the alloy is ferromagnetic (FM) at room temperatures, but it becomes paramagnetic (PM) when it transitions to the ordered B2 phase. This transformation occurs in the alloys with a Fe content ranging between $x$=0.5 and $x$=0.7 at aging temperatures $T_a$ below the critical temperature $T_c$ for the B2↔A2 phase transition, as illustrated in the structural (Figure 1b [30]) and magnetic (Figure 1c [19]) phase diagrams of this system.

The established view in the literature [19-29] is that the ordering transformation in $Fe_xAl_{1-x}$ is a single-phase process, where magnetism is progressively suppressed due to continuous atomic ordering, as schematically illustrated in Figure 1d for $Fe_{0.6}Al_{0.4}$.

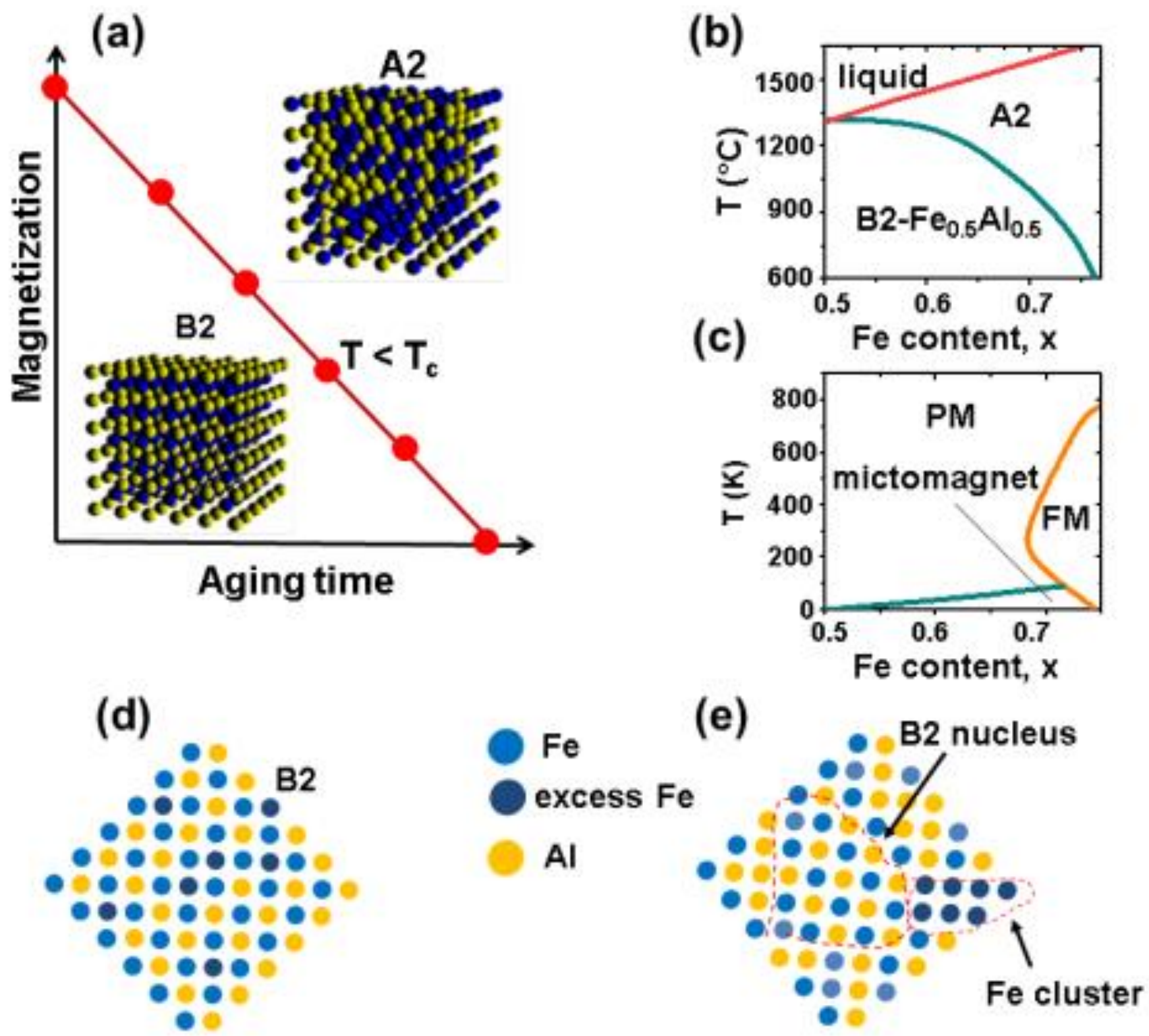

**Figure 1**: (a) Schematic plot of magnetization as a function of aging time in $Fe_xAl_{1-x}$ ($0.5<x<0.7$) at $T<T_c$. The insets show the atomic structure of the alloy in its two structural states. In the disordered A2 state, the alloy is ferromagnetic (FM), while it becomes paramagnetic (PM) in the ordered B2-$Fe_{0.5}Al_{0.5}$ state. (b-c) Structural (b) and magnetic (c) phase diagrams, which are adapted from Ref. [30] and Ref. [19], respectively. In the structural phase diagram, the A2↔B2 and solid↔liquid transitions are indicated by black and red lines, respectively. The magnetic phase diagram is plotted for the ordered B2-$Fe_xAl_{1-x}$ alloy, assuming continuous ordering of the alloy, as illustrated in (d). In the diagram, the PM↔FM and PM↔mictomagnet [4] transitions are indicated. (d-e) Schematic illustrations for two possible scenarios of ordering in a nonstoichiometric binary alloy. One of them (d) is continuous ordering at which excess atoms occupy sites randomly on the sublattice of another component. Another one (e) takes place through nucleation and growth of an ordered stoichiometric phase, with segregation and clustering of excess atoms.

In this model, excess Fe atoms randomly occupy the Al sublattice. This ordering mechanism is thermodynamically permissible, as it aligns with the Lifshitz criterion for second-order transformations in the CsCl-type lattice [31].

However, an alternative, first-order transformation pathway involving the nucleation and growth of a stoichiometric ordered phase is also plausible, particularly within a miscibility gap [32]. Such a heterogeneous process (Figure 1e) is often catalyzed by lattice defects [33-36] and, crucially, would lead to a fundamentally different microstructure. Indeed, growing evidence from studies on mechanical strengthening in Fe-Al-based alloys confirms that B2-phase precipitation within a disordered matrix is a viable mechanism [37-41]. Yet, the impact of this alternative ordering scenario on the system's magnetic and transport properties remains largely unexplored, creating a critical knowledge gap. Our work addresses this gap by directly investigating the high-temperature ordering kinetics and its consequences for magnetism. In our work, we have investigated the kinetics of chemical ordering in thin (50 nm) films of nonstoichiometric $Fe_xAl_{1-x}$ alloys ($0.5<x<0.7$) using methods of both structural and magnetic characterization. In contrast to previous studies [19-29], we found that, at sufficiently high $T_a>600$ °C but still lower than $T_c$, chemical ordering is accompanied by not a decrease in magnetization but rather by its increase. This improvement of the magnetism is also reflected in other properties such as the intensity of the magneto-optic Kerr effect (MOKE) and the giant anomalous Hall (AH) resistivity $\rho_{AH}$ found recently in Fe-Al alloys [5, 6, 8. 10, 29]. We have revealed, however, that high-temperature aging results in an even higher $\rho_{AH}$ than that in the as-grown sample, which has a higher magnetization. To understand these behaviors, we analyzed our samples using transmission electron microscopy (TEM), including the Electron

Spectroscopy Image (ESI) and selective area electron diffraction (SAED) modes. In these studies, we found that the ordering occurs through classical nucleation and growth of nuclei of the ordered B2-$Fe_{0.5}Al_{0.5}$ (B2) phase, so that the growing nuclei force excess Fe atoms to segregate into the surrounding matrix. This scenario for chemical ordering is schematically illustrated in Figure 1e. In fact, the enhancement of magnetization – we observe in our films at high $T_a$ – can reflect the clustering of excess Fe in the alloy. The clustering tendency is indicated in the $\rho_{AH}$ data we have obtained. Our experimental observations are supported by molecular dynamics (MD) simulations.

Thus, we demonstrate that the long-standing belief that chemical ordering invariably suppresses magnetism is not universal. With the prototypical $Fe_xAl_{1-x}$ system, we establish that high-temperature aging triggers an ordering mechanism via B2 discontinuous nucleation, enabling the enhancement of magnetization and related properties due to the formation of Fe-rich magnetic clusters.

# Methods

## Sample preparation

50 nm thick $Fe_xAl_{1-x}$ films were prepared with dc magnetron co-sputtering from individual high purity iron (Fe – 99.99 %) and aluminum (Al – 99.999 %) targets at room temperature. As substrates, we used 100 Si wafers coated with 100 nm of SiO2 and 50 nm of Si3N4 (Si/SiO2/Si3N4), fused quartz with a thickness of 0.8 mm, and commercially available 50 nm thick Si3N4 membranes. For film sputtering, an AJA 2200 multichamber system was employed at a basic pressure down to ~$6\times10^{-8}$ mbar, with introducing argon gas of 99.999% purity into the chamber using gas mass flow controller attached to the system. The base pressure in the system during sputtering was maintained in the range of 4.0 – 5.0 x $10^{-3}$ mbar. Film composition was varied by changing magnetron powers and thus sputtering rates of Fe and Al and checked by measuring glancing incidence X-ray reflectivity (XRR) with a Bruker D8 Discover x-ray diffractometer. To determine the composition, XRR curves were modeled and fitted to the experiment with employing the DIFFRAC.Leptos 7.04 (Bruker AXS) software; see Supplemental Information for greater detail. Thermal treatments at temperatures up to 800 °C were carried out in a high-vacuum chamber, with residual gas pressure of approximately $10^{-4}$ mbar. In addition, a rapid thermal annealing (RTA) system was used for short-term (down to ~10 s) heat treatments at $T_a$ up to 1000 °C. The RTA process involved the use of He or Ar gases that flowed through a heated sample at a rate of 10 l/min and a pressure of 1.5 kg/cm$^2$.

## Sample characterization

The phase transformations that occur in our films were investigated by examining both their structural and magnetic properties. We studied the crystalline structure of the films using TEM, which was operated at a LIBRA 200 MC apparatus (Carl Zeiss, Jena), operating at 200 kV. The apparatus was equipped with an OMEGA electron energy loss spectrometer, which allowed us to map the distribution of Fe and Al over the film surface in the ESI mode. In the TEM experiments, we determined the distribution in grain sizes as well as the LRO by analyzing SAED patterns. In the as-grown films, the average crystallite diameter was evaluated as 10.5±1.0 nm [26, 27]. The order degree was determined as $\mathrm{LRO} = \sqrt{(I_s/I_f)_{\mathrm{exp}}}/\sqrt{(I_s/I_f)_{\mathrm{id}}}$, [42] where $(I_s/I_f)_{\mathrm{exp(id)}}$ are ratios between intensities of a superstructure ($I_s$) and fundamental ($I_f$) diffraction peaks, which are found experimentally $[(I_s/I_f)_{\mathrm{exp}}]$ and taken from the calculations for the B2 structure $[(I_s/I_f)_{\mathrm{id}}]$ .

The magnetic properties were studied at room temperature with ferromagnetic resonance (FMR), MOKE magnetometry, and by measuring the AH resistivity. FMR data were collected with a Bruker EMX 10/12 spectrometer of electron paramagnetic resonance at a fixed frequency of 9.5 GHz. The detection of the resonance lines from the uniform precession modes allows for determining the saturation magnetization $4\pi M$ from the Kittel resonance equation. The used MOKE setup was home-built and based on a Faraday modulator technique. As a light source, a He-Ne laser ($\lambda$=633 nm, 5 mW, Thorlabs HRS015B) was employed. The MOKE intensity was measured as a function of magnetic field $H$ applied in the film plane to generate a MOKE hysteresis loop. The AH resistivity was measured using the standard four-probe method by applying a dc current of $I$=10 mA. Original data obtained by taking FMR spectroscopy and MOKE magnetometry are presented in Supplemental Information.

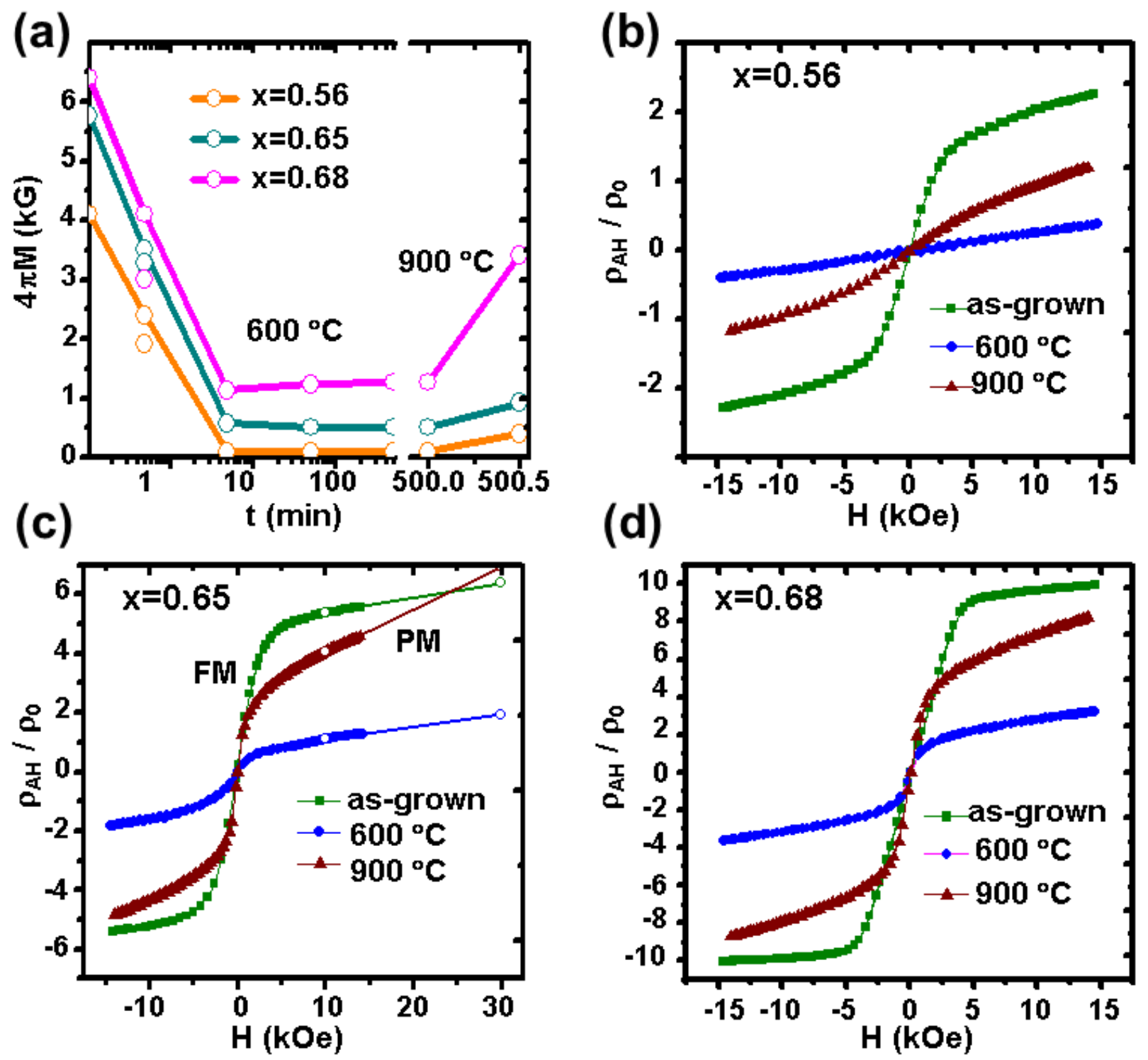

**Figure 2**: (a) Time evolution of magnetization $4\pi M$ in the films with different Fe content, which were aged at 600 °C and subsequently at 900 °C. The quantity of $4\pi M$ was measured with FMR by using Kittel resonance equations. (b-d) Room temperature AH resistivity $\rho_{AH}/\rho_0$ ($\rho_0$ is this quantity for pure Fe) versus applied magnetic field $H$ for the samples with $x$=0.56 (b), $x$=0.65 (c), $x$=0.68 (d) in the as-grown films as well as after film aging at 600 °C and 900 °C. There are fast (FM) and slow (PM) stages in the $\rho_{AH}$ evolution. The increase in the PM susceptibility after aging at 900 °C, results from a contribution of superparamagnetic clusters, which form owing to squeezing out the excess Fe from regions of the B2 phase.

## Modeling

MD simulations were performed using the LAMMPS [43] package to investigate the relaxation of a $Fe_xAl_{1-x}$ ($x$=0.6) crystal to the thermodynamic equilibrium at elevated temperatures. The initial atomic configuration was constructed based on the crystallographic data for the disordered A2 state, obtained from the CIF file (607483-ICSD) for the $Fe_{0.75}Al_{0.25}$ alloy. In this structure, a part of Fe atoms was substituted with Al in a random manner, in order to change the composition from $x$=0.75 to the desired one, $x$=0.6.

The crystal was generated by replicating the unit cell in three dimensions to form a 5×5×5 nm$^3$ a monocrystalline supercell consisting of different random realizations of the disordered structure from one unit cell to another one. Periodic boundary conditions were applied along all three Cartesian directions. Interatomic interactions were described using an Embedded-Atom Method (EAM) potential [44]. The resulting atomic configurations were visualized and analyzed using the OVITO software package [45].

## Results

### Enhancing the magnetism and AH resistivity

In our $Fe_xAl_{1-x}$ films, we have investigated how the magnetization $4\pi M$ determined using FMR as well as some related properties such as the MOKE intensity and $\rho_{AH}$, behave with increasing $T_a$. At sufficiently low $T_a \leq 600$ °C, the samples studied exhibit a decrease in $4\pi M$, down to vanishing room-temperature FMR responses in the samples with a sufficiently low Fe content, $x$<0.6. Basically, such a behavior is consistent with the previous data reported [19-29]. However, we observe the opposite trend at a higher $T_a$.

Figure 2 (a) shows $4\pi M$ versus aging time $t$ for the samples with different compositions – $x$=0.56, $x$=0.65, and $x$=0.68. The films were aged at 600 °C for 0.5, 5, 50, and 500 min. Also, the samples aged at 600 °C for 500 min were subsequently aged at 900 °C for 0.5 min in the RTA system. We see that, after RTA aging at 900 °C, the samples with such compositions exhibit a significant increase in magnetization at room temperature. Quantitatively, the same trend was observed in the specimens aged at 800 °C for 1 hour in a high-vacuum chamber. This behavior is in contrast to what we observe at a lower $T_a \leq 600$ °C. We note that the Fe content of $x$=0.56 was lowest, at which we detected the reentrance of FMR response after elevation of $T_a$ up to 900 °C. The original FMR data for all compositions studied are presented in Supplemental Information.

Table 1: Comparison of RTA and longtime (1 hour) high-vacuum thermal annealing on MOKE; see Supplemental Information for the original data. The measured MOKE intensity is a difference in Kerr rotation in saturated positive and negative applied magnetic fields, $H$. In brackets (sample with $x$=0.56), the MOKE intensity is indicated, which was measured by taking a polar MOKE at $H$=±5 kOe. The film composition was changed by varying the magnetron powers indicated in the first column by numbers in units of W. $2\theta_c$ is the double total external reflection edge, which is related to the average dielectric susceptibility $\langle\varepsilon\rangle$ across the film as $\cos^2\theta_c=\mathrm{Re}\langle\varepsilon\rangle$ [47].

| | | | MOKE intensity, mrad | | | |
|---|---|---|---|---|---|---|
| sample | $2\theta_c$, degr. | x | as-grown | 600 °C 100 s / 1 hour | 800 °C 100 s / 1 hour | 900 °C 30 s |
| Fe(110)-Al(150) | 0.67 | 0.68 | 1.4 | 0.25 / 0.2 | 0.5 / 0.4 | 0.7 |
| Fe(100)-Al(150) | 0.66 | 0.65 | 0.8 | 0.1 / 0.05 | 0.15 / 0.1 | 0.3 |
| Fe(85)-Al(150) | 0.64 | 0.56 | 0.2 (1.0) | <0.05 | <0.05 | <0.05 (0.05) |

Here, we also emphasize that we do not observe the oxidation effects [46] on magnetization of the samples aged at high $T_a$ up to 900 °C (at least, at short annealing times ~ 100 s), We find, for example, that RTA under He flow at the air atmosphere and high-vacuum annealing induce the same changes in magnetization. The MOKE intensity measured for the as-grown samples and after RTA and high-vacuum thermal annealing is illustrated in Table 1. The original MOKE data obtained (MOKE hysteresis loops) are presented in Supplemental Information for all compositions studied.

It is of special interest to study the AH resistivity of $Fe_xAl_{1-x}$ [5, 6, 8, 10, 29]. In such Fe-based alloys, due to their half-metallicity and thus high spin polarization [1, 2], the AH effect can be called as the *giant* AH effect, which can be even much larger than that in pure Fe ($\rho_0$=0.18 μΩ cm [10]). The empirical relationship between the AH resistivity and magnetization can be expressed as $\rho_{AH} = R_s 4\pi M_z$, where $R_s$ is the conventional notation for the AH coefficient and $4\pi M_z$ is the out-of-plane component of magnetization. In our case, the studies of $\rho_{AH}$ enable a new kind of information about the ordering tendencies in the samples under study. Figure 2 (b-d) shows our results of measuring $\rho_{AH} = V_H L/I$ $\rho_{AH}=V_H L/I$ ($V_H$ is the measured Hall voltage, $L$ is the film thickness, and $I$ is the current passed through the film) as a function of applied magnetic field $H$. In the plots, the quantity of $\rho_{AH}$ is normalized by $\rho_0$. The dependencies obtained are shown for the samples in their as-prepared state as well as after aging at 600 °C and 900 °C. We find that both as-grown and aged films exhibit the two-stage behavior of the $\rho_{AH}$ evolution with increasing $H$. The occurrence of fast and slow stages in the $\rho_{AH}$–versus-$H$ dependence can be associated with the FM and PM phases, respectively. It is striking that there is a significant ehancement of the PM susceptibilty after aging at 900 °C by comparison that at 600 °C, as indicated by tilted thin lines. We see that, at sufficiently high $H$>20 kOe, the quantity of $\rho_{AH}$ in the sample aged at 900 °C can even exceed that for the as-grown film, whose magnetization remains much higher.

## LRO determination

To elucidate the relationship between the magnetism and LRO in our films, we determined the intensity of diffraction rings in the SAED patterns as functions of the aging time ($t$) at the same aging temperatures, 600 °C and 900 °C. It is found that the increase in magnetization and related properties we observe is accompanied by a further increase in diffraction intensity from the B2 phase, which is present even in the as-grown films [26, 27]. Generally, chemical ordering in a magnetic alloy has to suppress the ferromagnetism because of reducing the number of nearest Fe neighbors in the atomic lattice [48].

Figure 3a shows the time evolution of the ratio between intensities of the strongest B2 (100) and fundamental (110) diffraction peaks, $I_{100}/I_{110}$, the square root of which is proportional to LRO [42]. A representative SAED pattern [26, 27] of the aged samples, with indication of the 100 (highlighted in red) and 110 diffraction rings, is presented in Figure 3b. In greater detail, Figure 3 (c-d) depicts the SAED intensity versus the scattering vector $q$=2sinθ/λ with θ and λ being the electron diffraction angle and electron wavelength, respectively. As seen from Figure 3c, up to three B2 peaks – 100, 111, 210 – are detectable in the used SAED mode. Positions of the diffraction maxima are at $q_{hkl} = \sqrt{h^2 + l^2 + k^2}/a$, where $h$, $k$, $l$ are Miller indices and $a$ is the interplane distance. For the B2 phase, the latter quantity is 0.287 nm, 0.291 nm and 0.288 nm in the as-grown sample, after aging at 600 °C, and after aging at 900 °C, respectively. We also note that, as shown in Figure 3d, a new diffraction ring is detectable at $q$=4.2 nm$^{-1}$ after RTA aging at 900 °C, which can be attributed to an oxide phase such as $Fe_2O_3$ or $Al_2O_3$. As there is no its effect on the magnetization (Table 1), one may suppose that this peak occurs from the oxide passivation layer, which forms on the sample surface under a heat treatment.

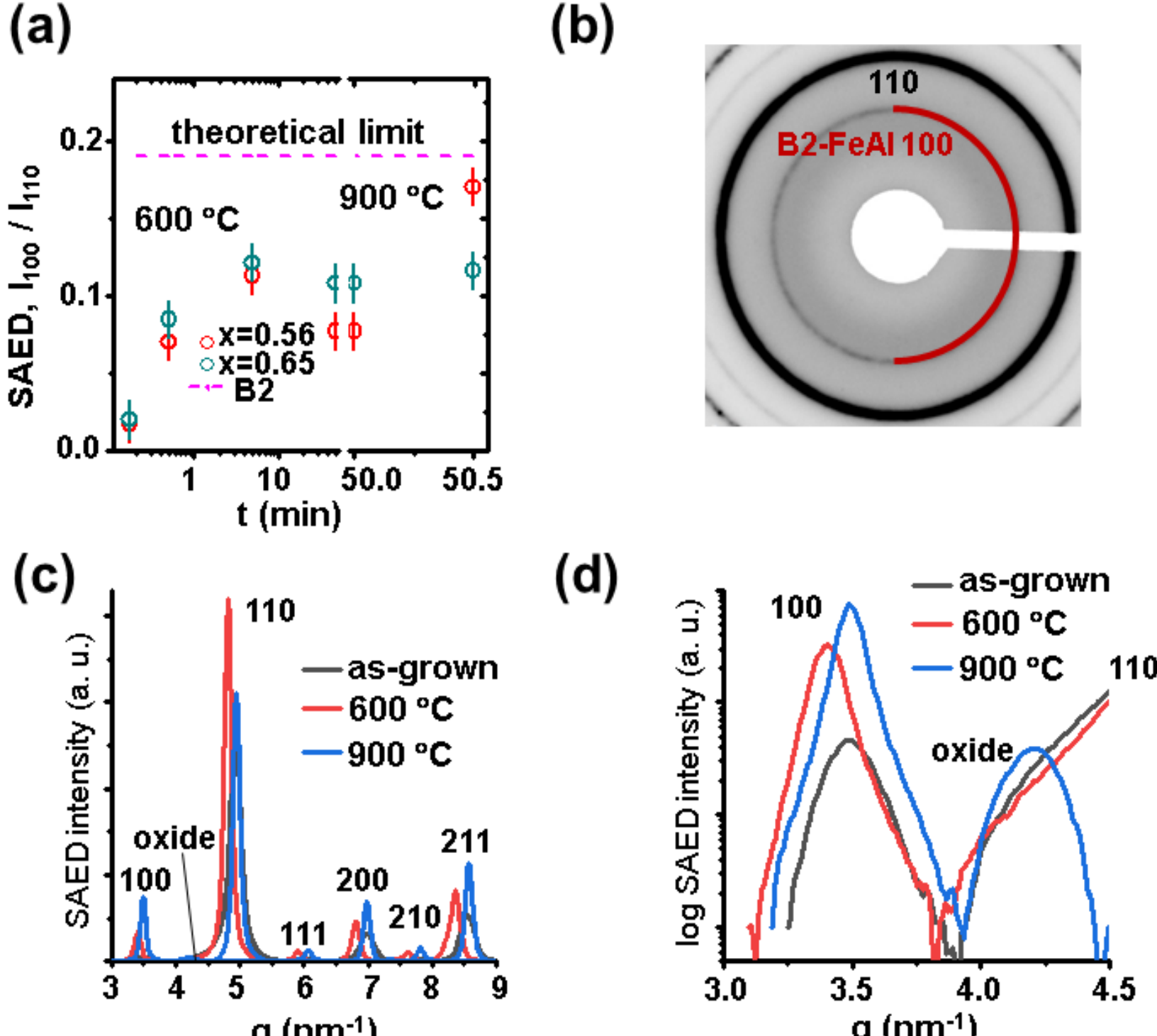

**Figure 3**: (a) Time evolution of the quantity of $I_{100}/I_{110} \propto$ LRO$^2$, where $I_{100}$ and $I_{110}$ are intensities of the strongest B2 (100) and fundamental (110) diffraction peaks. The dependences obtained by determining the intensities of diffraction rings in the SAED patterns. The horizontal dashed line (theoretical limit) is the calculated ($I_{100}/I_{110}$) for the B2 structure [49]. (b) A representative SAED pattern, which indicates the strongest B2 (100) and fundamental (110) diffraction peaks. (c) SAED intensity versus scattering vector q in the as-grown sample with $x$=0.56 and after its aging at elevated $T_a$, with indication up to three B2 diffraction peaks – 100, 111, and 210. (d) The SAED intensity in the logarithmic scale, with indication the strongest B2 diffraction peak (100) as well as the diffraction peak from an oxide layer on the sample surface.

From the plotting in Figure 3, we see that after aging at 900 °C the superstructure peak intensity becomes larger than that at 600°C, at which the magnetization remains still small after heat treatment (Figure 2a). Moreover, it is even can be close to its theoretical limit [49] indicated in Figure 3a (dashed horizontal line). Thus, we find that, in our $Fe_xAl_{1-x}$ sample aged at high $T_a$ (but still lower $T_c$), there is an inconsistence between the evolutions of magnetization and LRO.

## B2 nucleation: Experiment and modeling

To understand a reason(s) for the observed (1) inconsistency between the magnetism and LRO and (2) enhancement of magnetism at high-temperature aging, the film crystalline structure was examined with TEM, including the ESI mode. Figures 4a and 4b depict TEM bright field overview micrographs of the sample with $x$=0.56 after aging at 600 °C and subsequently at 900 °C. We note that the SAED pattern corresponding to the image for the sample after aging at 900 °C shows a high B2 diffraction intensity, as illustrated in Figure 3a. Both TEM overview micrographs indicate that the film microstructure is inhomogeneous to contain bright regions embedded in the crystal structure of the alloy. As $T_a$ increases from 600 °C to 900 °C, we observe enlargement of the bright regions up to submicron dimensions, as shown in Figure 4b.

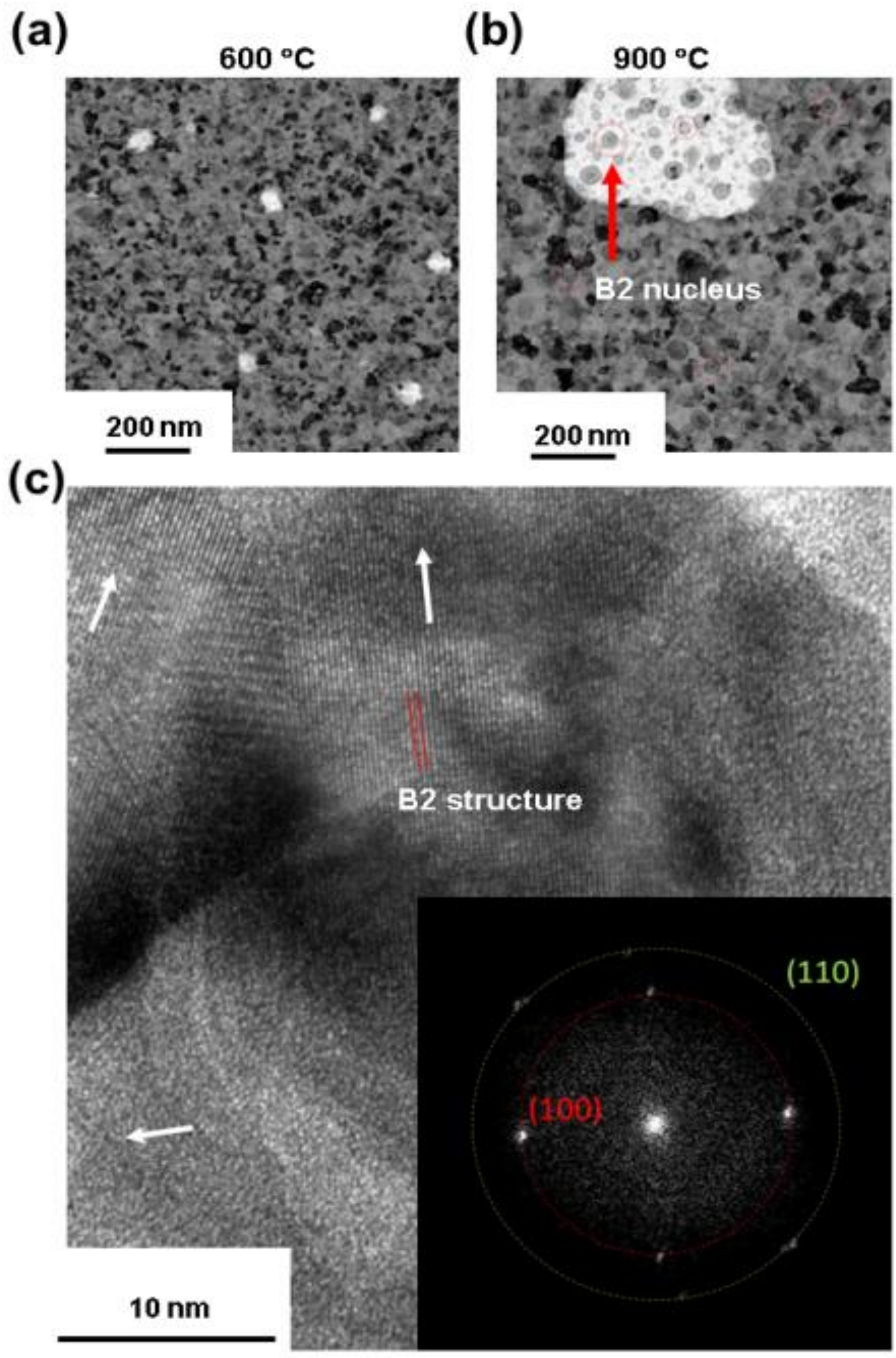

**Figure 4**: (a-b) TEM bright field overview micrographs of the sample with *x*=0.56 aged at 600 °C (a) and 900 °C (b). The bright regions in the images consist of assemblies of B2 nuclei, which are highlighted with red contours. (c) HRTEM micrograph of the morphological structure, where a few B2 nuclei are identifiable. The inset is the Fast Fourier Transform of the HRTEM micrograph, which indicates the formation of the B2 structure through the appearance of the 100 reflection.

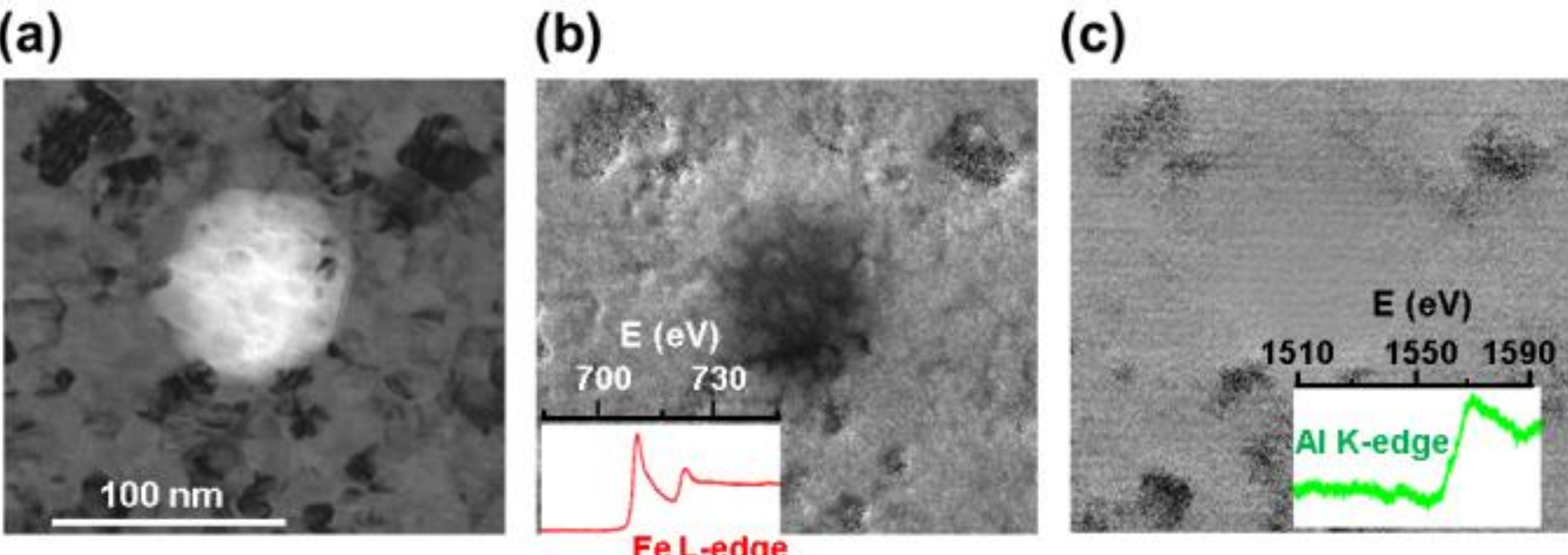

**Figure 5**: (a) Zero loss bright field TEM micrograph of a film location, which contains a zone with a high concentration of B2 nuclei (bright region in the center). (b-c) ESI maps of Fe and Al concentration distribution near the Fe L (b) and Al K (c) absorption edges, which are taken inside the bright region. The insets show the corresponding electron energy loss spectra averaged through the scanned area.

Examining the bright regions, like that seen in top of Figure 4b, has revealed that they contain the assemblies of well-shaped crystals with diameters of 10-20 nm (highlighted with red contours). Figure 4c shows a high-resolution TEM image and its Fast Fourier Transformation (FFT) in the inset, where a few such nanocrystals can be identified (indicated by arrows) through the effect of the periodic atomic arrangement in the alloy. The FFT we show allowed us to reveal the 100 reflection from the B2 structure in addition to the 110 fundamental one.

However, the question about identifying the ordering scenario (Figure 1 d, e) still remains. To find an answer to that question, ESI maps were collected inside of a one of the bright regions seen in Figure 4a, which contain a higher concentration of B2 nuclei than the environment does. In Figure 5a, we show a TEM micrograph of a film location, which contains this region. Figures 5b and 5c show the distributions of Fe and Al concentrations over the scanned location, respectively. The elemental maps were obtained near the Fe L and Al K absorption edges using a 10-eV spectrometer slit width during 20-minute exposure for each signal and background image. The one-window contrast evaluation method was used to estimate the background signal, which allowed for a qualitative evaluation of the chemical distribution of elements in the alloy.

We have found that the Fe concentration within the bright zone is lower than that of the surrounding matrix (Figure 5b), while the Al concentration remains uniform throughout the film area (Figure 5c). Thus, we conclude that achieving a higher LRO corresponds to a higher concentration of B2 nuclei in the aged alloy, which is accompanied by squeezing out the excess Fe into the surrounding matrix.

Figure 6 shows the atomic structures in the $Fe_{0.6}Al_{0.4}$ crystal, which were simulated with the MD method at 1100 °C with different aging times – 10 ns, 40 ns, and 60 ns. During the early stages of ordering occurring through thermal vacancies in the crystal lattice, the B2 nucleation was observed. For example, we see (1) the formation of a B2 nucleus in the right upper corner of the "40 ns" pattern and (2) segregation and clustering of Fe (blue tones), as pointed out by arrows. It is interesting to note that, in such a temperature regime, the ordering process is activated within the nanosecond timescale. Indeed, there is a pronounced difference in atomic arrangements of the "10 ns" and "40 ns" patterns, while the "40 ns" and "60 ns" patterns are already close each to other. Such an ordering rate is compatible with analytical simulations performed earlier [50] by combining the Landau theory for second-order phase transitions [51] and the Metiu, Kitahara, and Ross equation for LRO relaxation [52].

## Discussion

Based on the FMR/MOKE, TEM/SAED/ESI, $\rho_{AH}$-versus-$H$ data, and MD simulations we find that, after aging at 900 °C, the alloy appears to be compositionally nonuniform, consisting of two regions. One of them in this dual-phase hybrid is a nearly pure B2 phase. Indeed, the 100 diffraction ring intensity becomes close (Figure 3a) to that calculated for the B2 structure, $I_{100}/I_{110}$=0.19 [49].

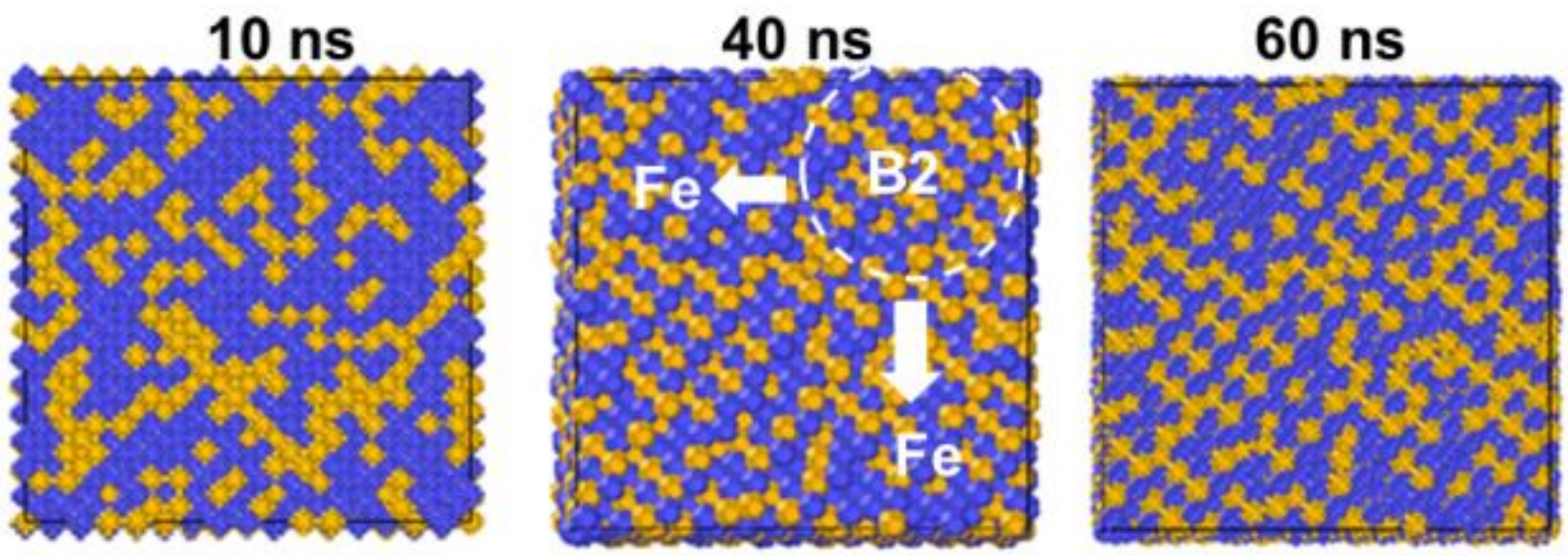

**Figure 6**: MD simulations of the B2 nucleation in the $Fe_{0.6}Al_{0.4}$ alloy aged for different times at 1100 °C. Blue and yellow tones are Fe and Al atoms, respectively. The formation of the B2 nucleus, indicated by the dashed contour, occurs after aging for 40 ns. As a result of this process, excess Fe atoms are expelled to other locations, as indicated by arrows.

Another phase is Fe-enriched precipitates embedded in the B2 matrix. It is highly likely that the clustering of excess iron is responsible for the observed improvement in the magnetism (Figure 2a). By contrast, we observe mostly a decrease in magnetization and related properties at 600 °C, and so, segregation of Fe and its precipitation are not effective in such a heat treatment regime.

We should also note that the ordering scenario through nucleation and growth of the B2-$Fe_{0.5}Al_{0.5}$ phase (Figure 1e) in the nonstoichiometric alloy is incompatible with the magnetic phase diagram adopted for $Fe_xAl_{1-x}$ (Figure 1c). Although this diagram is derived from experimental data, it can be interpreted as showing magnetic phases and their transitions in a single-phase B2-$Fe_xAl_{1-x}$ alloy (Figure 1d). However, Fe precipitates in the ordered alloy have to increase the critical temperatures for magnetic transitions, which are displayed in the current magnetic diagram to be not higher than a few tens of Kelvin.

Finally, we would like to emphasize that high-temperature aging (900 °C) causes the further ordering in the alloy with massive B2 formation, in which magnetization is very small ($4\pi M<0.1$ kG) [19, 22, 24]. Therefore, the explanation of the increase in $\rho_{AH}$ we observe (Figure 2b-d) can be found by considering the contributions from superparamagnetic clusters [27], which form owing to squeezing out the excess iron by growing B2 nuclei. Sizes of these clusters can be roughly evaluated from the following relationship between magnetization $M$ and applied magnetic field $H$ in a thin PM film:

$$M = \frac{C_J}{T+(4\pi/3)C_J} H, \quad (1)$$

where $T$ is the ambient temperature, $C_J=N\mu^2/3k_B$ is the Curie constant, $N$ is the concentration of superparamagnetic clusters, μ is the total magnetic moment of a cluster, and $k_B$ is the Boltzmann constant. Equation (1) is valid if $\mu H<<k_BT$. It follows from this condition that the cluster can contain at most ~10 magnetic atoms at $T$ close to room temperatures, at which $\rho_{AH}$ was measured. It here is interesting that, at sufficiently high $H>20$ kOe, the contribution of the clusters into $\rho_{AH}$ can even exceed that in the as-prepared alloy, whose magnetization is higher.

## Conclusion

In summary, we have investigated the kinetics of chemical ordering in nonstoichiometric $Fe_xAl_{1-x}$ ($0.5<x<0.7$) thin-film alloys and uncovered a non-monotonic relationship between the magnetism and aging temperature. While annealing at lower temperatures ($T_a\leq600$ °C) leads to the well-established decrease in magnetization due to continuous atomic ordering, high-temperature annealing ($T_a=900$ °C) triggers an opposite trend, enhancing the magnetization and related properties such as the anomalous Hall resistivity.

The key to understanding this behavior lies in identifying a change in the fundamental ordering mechanism, which is the nucleation and growth of the stoichiometric ordered (B2) phase, a first-order transformation pathway. Crucially, this process forces the segregation of excess Fe into the surrounding matrix, as directly evidenced by our TEM/ESI analysis. The observed magnetization enhancement is a direct consequence of the formation of Fe-enriched clusters within the ordered B2 matrix. This qualitative model successfully explains the distinctive field evolution of the anomalous Hall resistivity. Our conclusions are further supported by molecular dynamics simulations of the $Fe_{0.6}Al_{0.4}$ alloy.

Thus, beyond the previously documented continuous ordering, Fe-Al alloys can access a competing ordering pathway via phase separation. This work clarifies the complex interplay between structural order and magnetism in this prototypical system and provides a revised framework for controlling functional properties in ordering alloys. We finally note that after their high-temperature treatments such alloys demonstrate excellent magnetic and electronic properties even with no sintering oxygen-containing solid solutions [46].

## Acknowledgements

The authors express their gratitude to Dr. M. Bohra (Mahindra University, Hyderabad, India) for fruitful discussions. This work was supported by Ministry of Science and Higher Education of the Russian Federation (State Contract FFUF-2024-0021). MD modeling was performed using the resources of the "Lobachevsky" computing cluster of the Lobachevsky State University.

**Supplemental Information** Raw Data: X-ray reflectivity (XRR) analysis, Ferromagnetic resonance (FMR) spectroscopy, and Magneto-optical Kerr effect (MOKE) magnetometry

# Enhancing the magnetism and giant anomalous Hall effect in thin Fe-Al films via B2 nanophase growth

D. A. Tatarskiy, A. A. Nazarov, Y. M. Kuznetsov, A. V. Zdoroveyshchev, I. Y. Pashenkin, P. A. Yunin, S. A. Churin, E. S. Demidov, M. V. Sapozhnikov, N. I. Polushkin

## Supplemental Information: Original data

### A. XRR curves for determining the composition

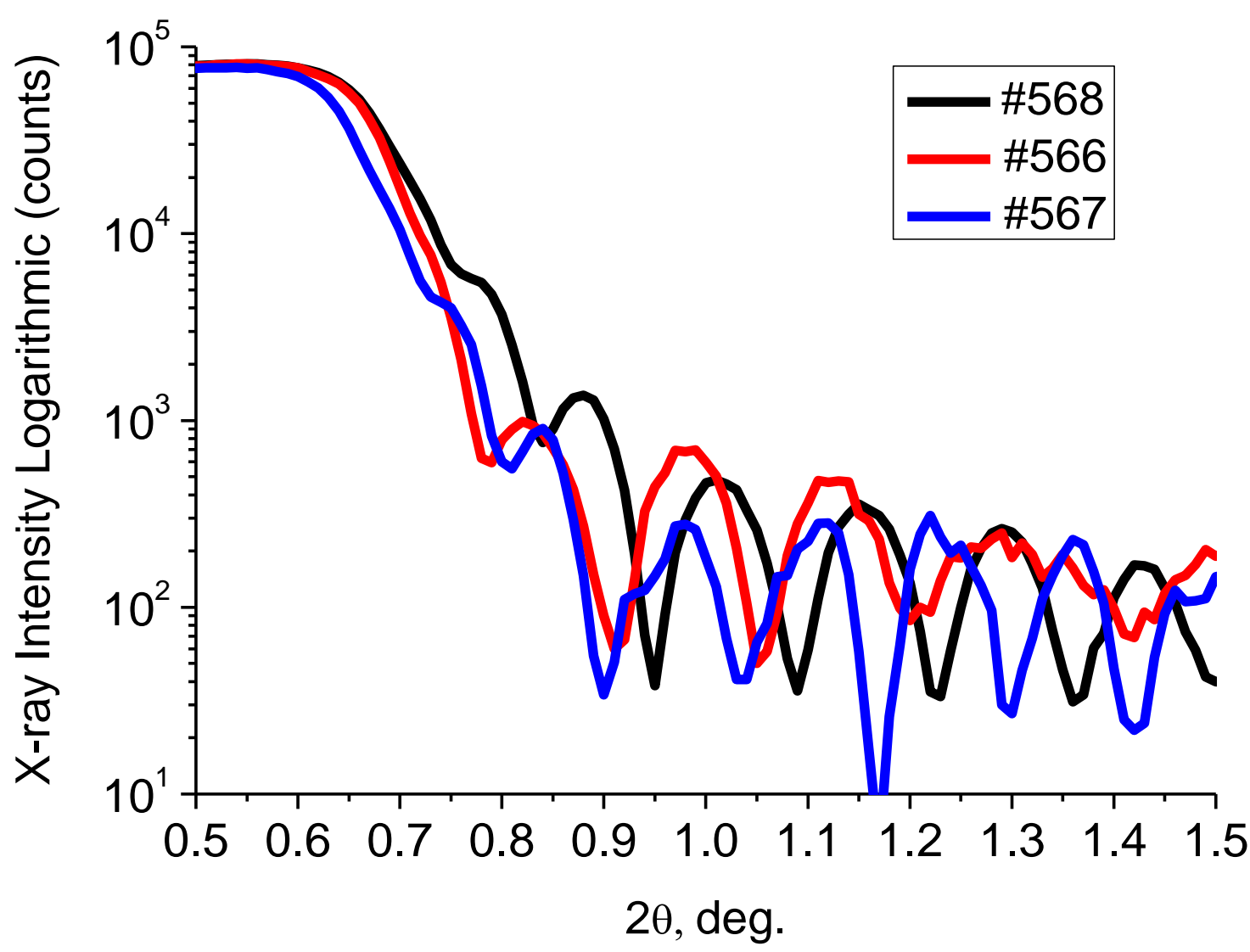

**Figure S1**: X-ray reflectivity (XRR) at the Cu $K_\alpha$ edge ($\lambda$=0.154 nm) versus the double angle of grazing incidence for three as-prepared films of the $Fe_xAl_{1-x}$ alloy sputtered on Si(500 μm)/SiO2(100 nm) /Si3N4(50 nm) substrates. The samples have different magnetizations measured by FMR and MOKE magnetometry. The composition, *x*, was determined using two methods. One way was to measure the edge of total external reflection as the maximum of the angle derivative of XRR. Another way was modeling XRR curves and their fitting to the experimental ones shown in this plot using DIFFRAC.Leptos 7.04 (Bruker AXS) software [A.Ulyanenkov, LEPTOS: A universal software for X-ray reflectivity and diffraction, Advances in Computational Methods for X-Ray and Neutron Optics, 5536 (2004) 1-15].

Fitting of interference fringes in XRR curves allowed for determining precisely the atomic density of a sample, $\rho_s$. The composition value was found as $x = (\rho_s - \rho_{Al})/(\rho_{Fe} - \rho_{Al})$, where $\rho_{Al}$ =2.7 g/cm$^3$ and $\rho_{Fe}$=7.9 g/cm$^3$ are the atomic densities of Al and Fe bulks. The results of the measurements are summarized are summarized as follows:

**Table S1:** $2\theta_c$ is the double total external reflection edge and $\rho_s$ are measured atomic densities of the surface oxide, Fe-Al films, and $SiO_2$ substrate. The thicknesses and surface roughness determined by fitting the modeled layered system to the experimental XRR curves are given as well.

- **#567 ($2\theta_c$=0.64°)**

| layer | Thickness, nm | Roughness, nm | ρs, g/cm$^3$ |
|---|---|---|---|
| oxide | 5.74 | 1.95 | 2.73 |
| Fe-Al | 56.2 | 2.41 | 5.58 |
| Substrate SiO2 | | 1.04 | 2.33 |

- **#566 ($2\theta_c$=0.66°)**

| layer | Thickness, nm | Roughness, nm | ρs, g/cm$^3$ |
|---|---|---|---|
| oxide | 7.2 | 4.7 | 5.48 |
| Fe-Al | 43.6 | 1.5 | 5.97 |
| Substrate SiO2 | | 1.22 | 2.33 |

- **568 ($2\theta_c$=0.67°)**

| layer | Thickness, nm | Roughness, nm | ρs, g/cm$^3$ |
|---|---|---|---|
| oxide | 5.5 | 2.0 | 3.0 |
| Fe-Al | 53.6 | 2.2 | 6.19 |
| Substrate SiO2 | | 1.1 | 2.33 |

# B. FMR spectroscopy

- **56 at. % Fe (#567)**

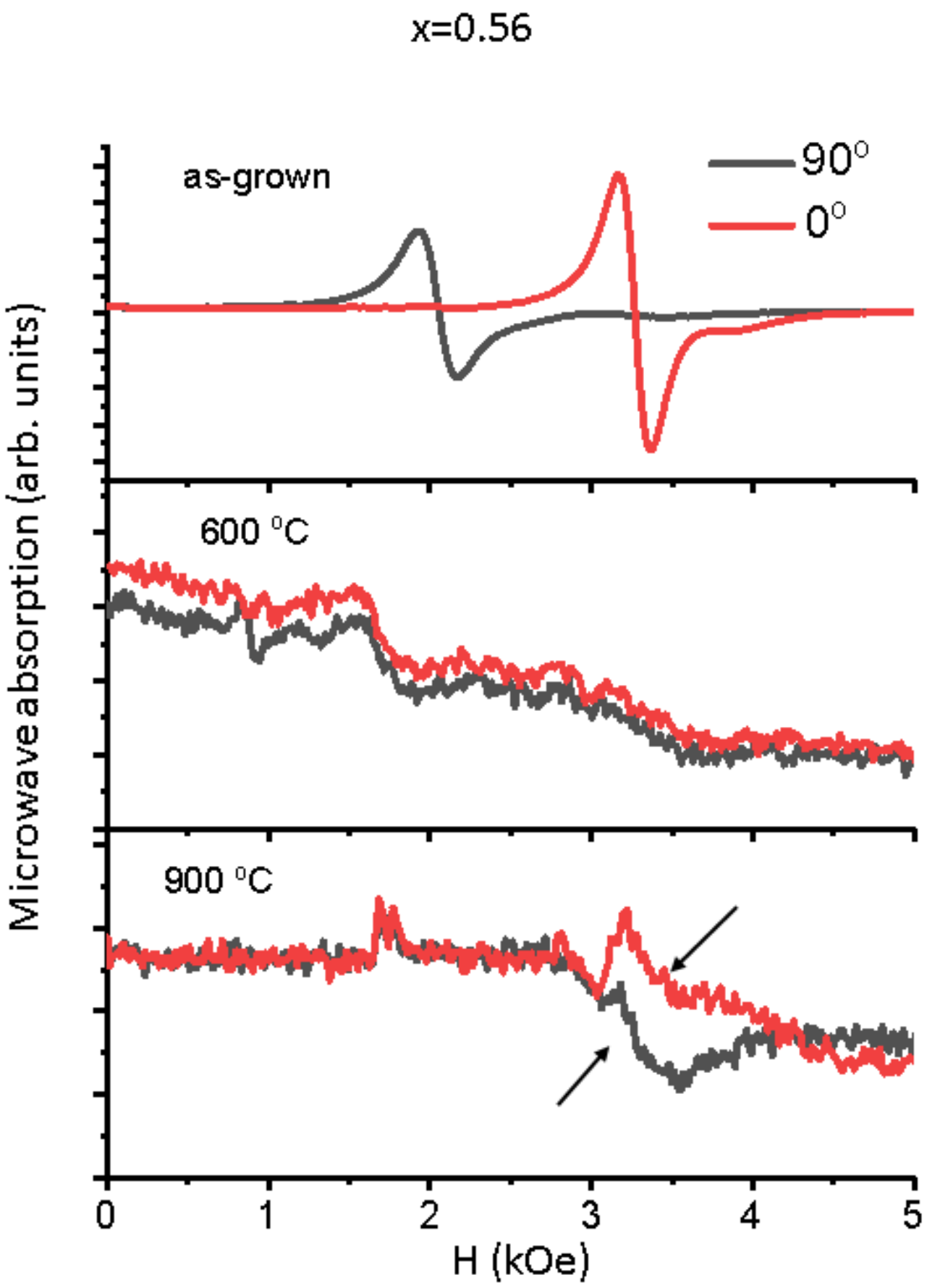

Figure S2: FMR spectra obtained at room temperature for the as-grown sample with x=0.56 and after its aging at elevated $T_a$. The spectra are shown for two geometries of an applied static field $H$: (1) the field is parallel to the film plane (90° ) and (2) the field is normal to the film plane (0°). The reentrance of the FMR response is observable after high-temperature aging at 900 °C (indicated by arrows).

- **65 at. % Fe (#566)**

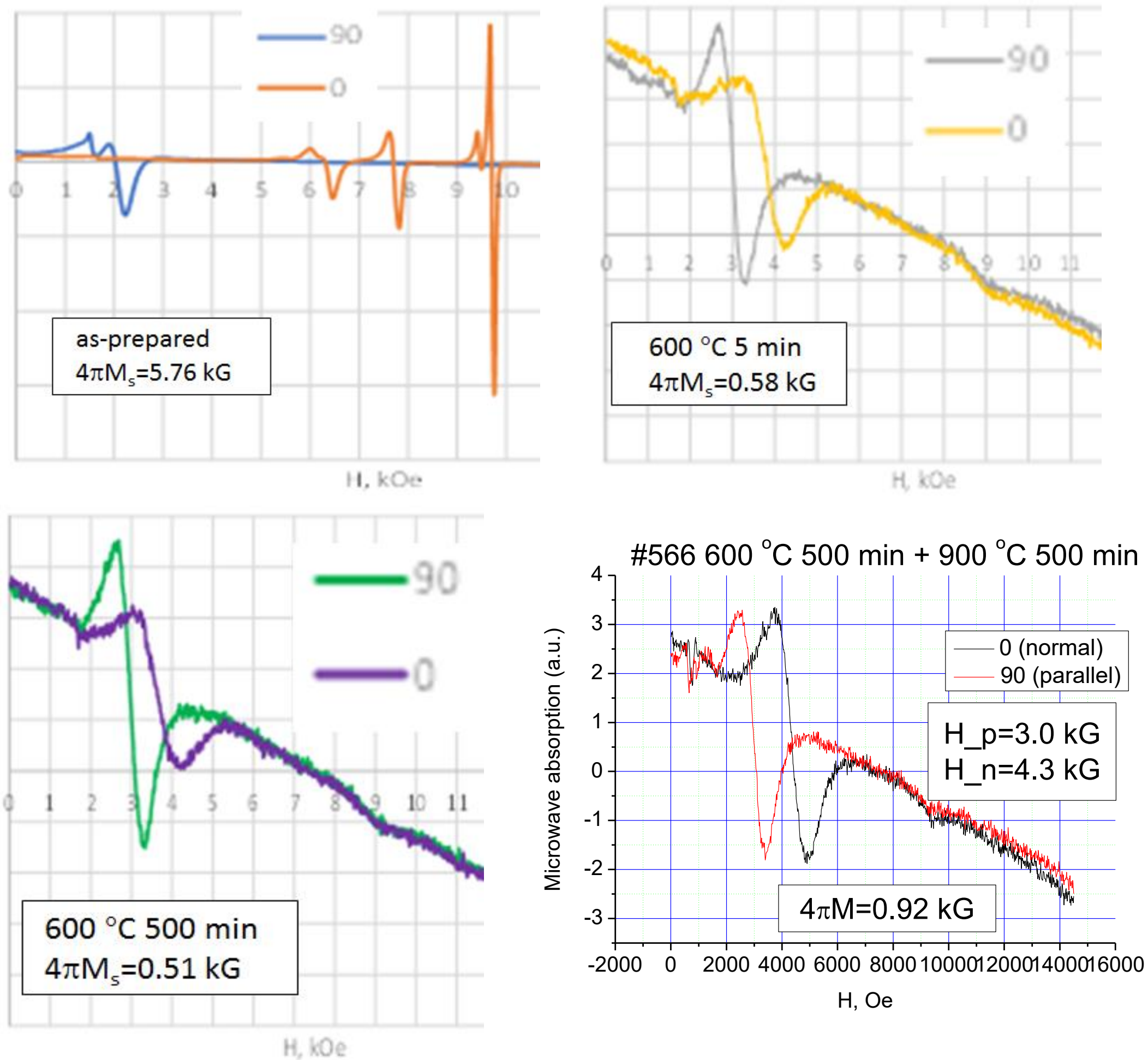

**Figure S3**: FMR spectra obtained at room temperature for the as-prepared, and aged sample with a composition of $x$=0.65. The spectra are shown for two geometries of an applied static field $H$: (1) the field is parallel to the film plane (noted as "90") and (2) the field is normal to the field plane (noted as "0"). The as-prepared film is not uniform

across its thickness.

- **68at. % Fe (#568)**

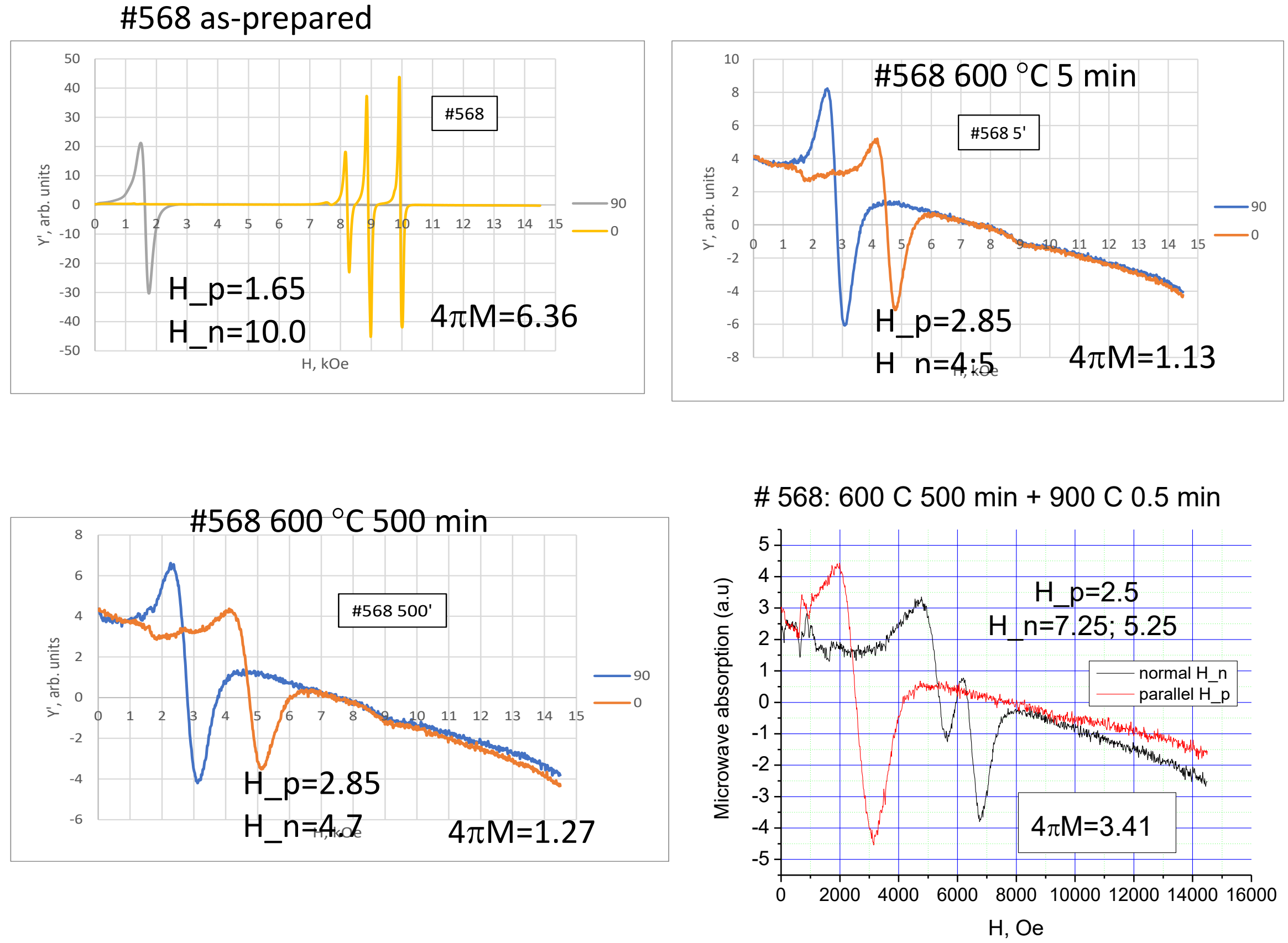

**Figure S4**: FMR spectra obtained at room temperature for the as-prepared, and aged sample with a composition of $x$=0.68. The spectra are shown for two geometries of an applied static field $H$: (1) the field is parallel (H_p) to the film plane and (2) the field is normal (H_n) to the field plane. The resonance fields are given in units of kOe and kG, respectively.

# C. MOKE magnetometry

**Figure S5**: Longitudinal/polar MOKE magnetometry of the as-prepared, aged, and quenched samples with 68, 65, and 56 atomic percent of Fe. The MOKE intensity is indicated in the plots.

- **68 at. % Fe (#568)**

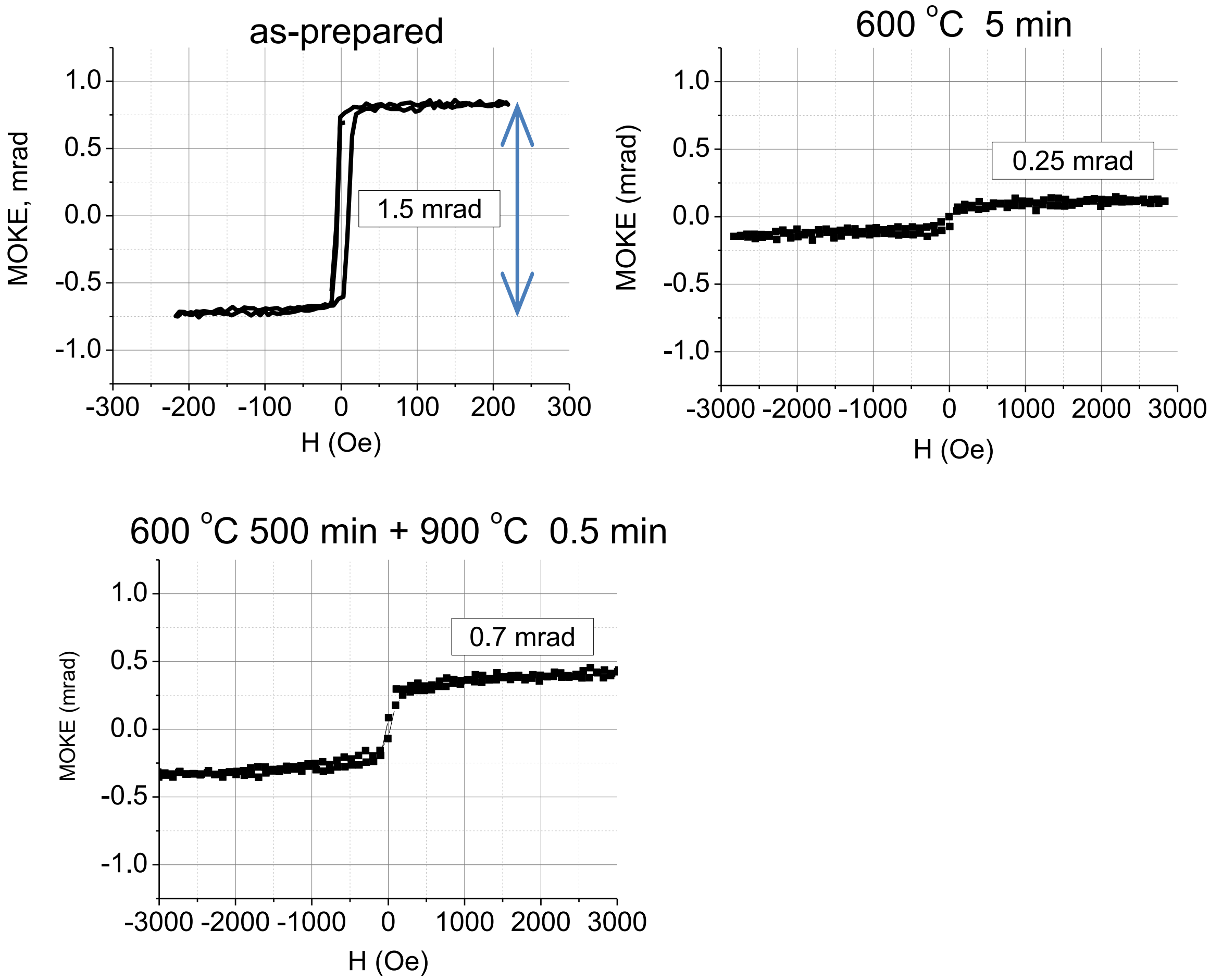

- **65 at. % Fe (#566)**

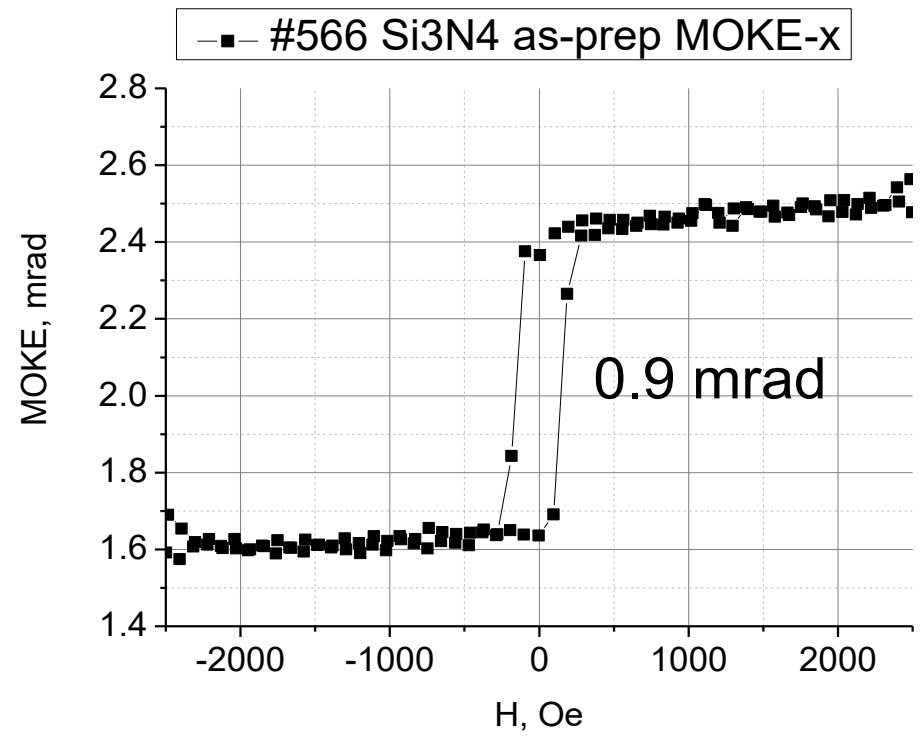

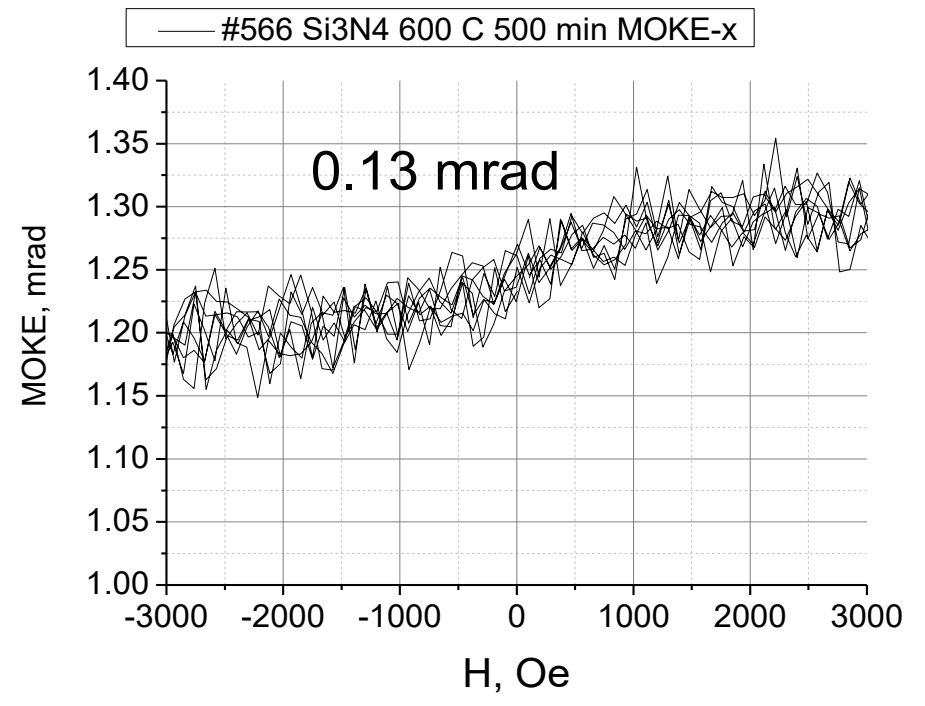

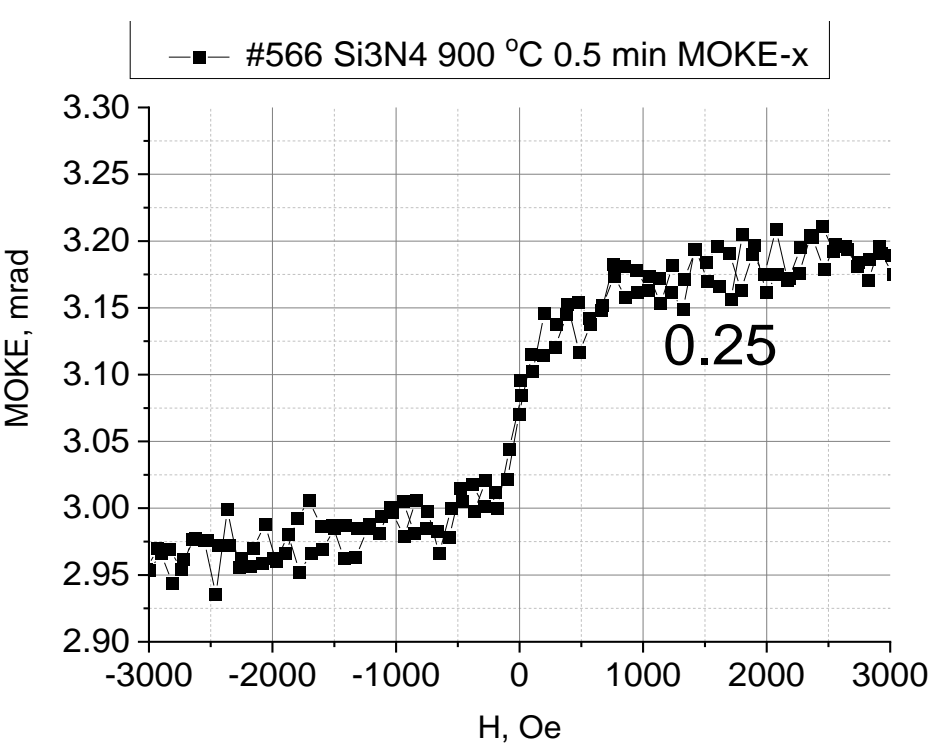

- **56 at. % Fe (#567)**

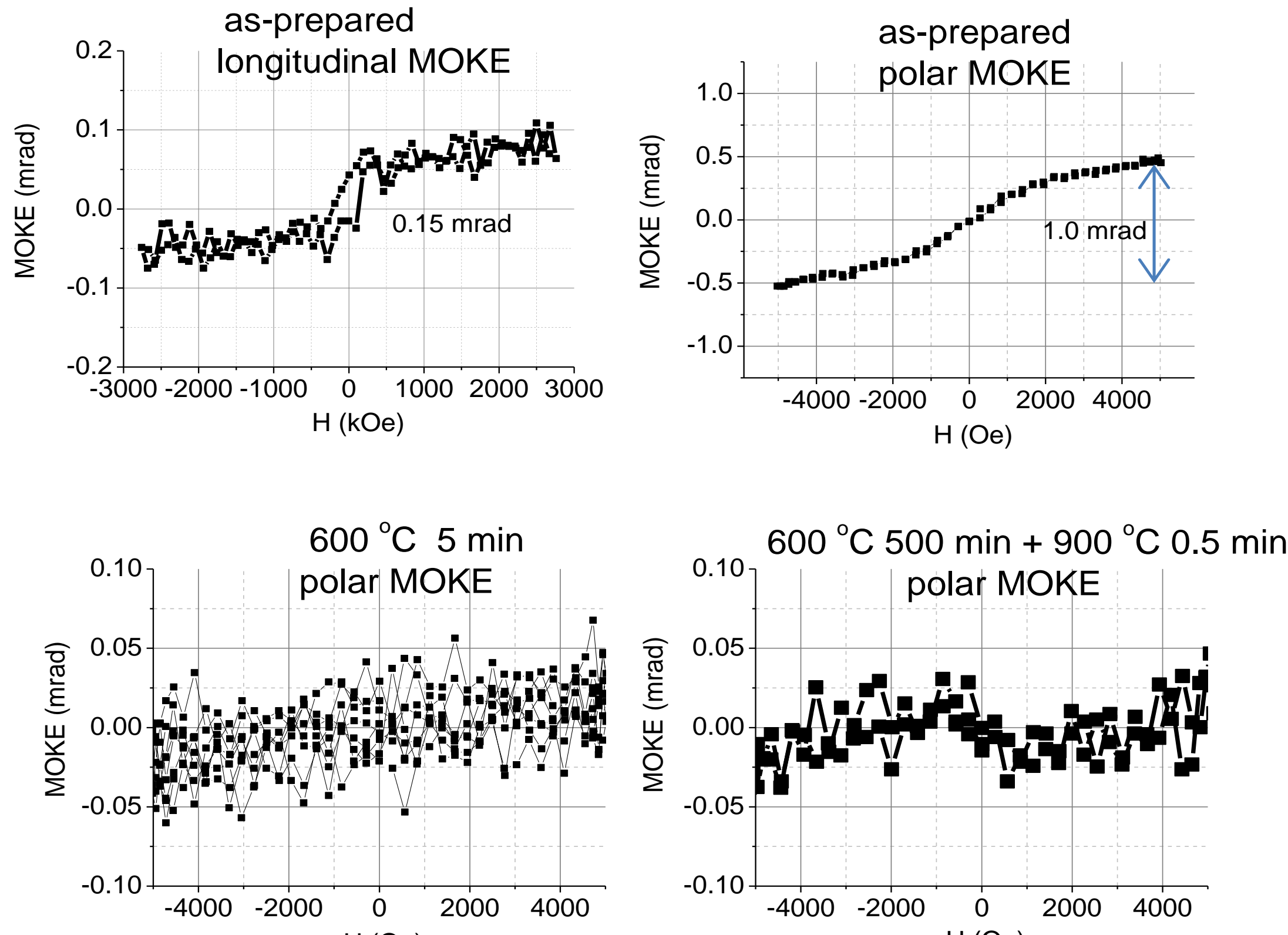